\documentclass[11pt,twoside]{article}
\usepackage{asp2010}

\resetcounters

\begin{document}

\title{Resolved Stars in M83 Based on HST/WFC3 Early Release Science Observations}
\author{Hwihyun~Kim$^1$, Bradley~C.~Whitmore$^2$, and Rogier~A.~Windhorst$^1$
\affil{$^1$School of Earth and Space Exploration, Arizona
  State University, Tempe, AZ 85287-1404, USA}
\affil{$^2$Space Telescope Science Institute, Baltimore, MD 21218}
}

\begin{abstract}
We present a multi-wavelength photometric study of individual stars in
M83 based on observations taken as part of the WFC3 Early Release
Science (ERS) program.  The central region of M83 has been imaged in
seven broad-band filters to obtain multi-wavelength coverage from the
near-UV to near-IR.  We use four filters--F336W, F438W, F555W, and
F814W--to measure the $UBVI$ photometric colors and intrinsic
luminosity for $\sim$10,000 stars.  These measurements are used to
determine the recent ($<$ 1 Gyr) star formation history of M83.  We
selected 50 regions in the spiral arm and the inter-arm area of M83
and categorize them based on their H$\alpha$ morphology.  To determine
ages of stars in each region, the color-magnitude diagrams (CMD) and
the color-color diagrams were used with the Padova isochrones with a
metallicity of Z=0.03 (1.5 Z$\odot$).  We correct the foreground and
internal extinction of individual stars using the reddening-free Q
parameter and improve our age estimates of stars from the isochrone
fitting on the CMDs.  Comparisons of stellar ages and cluster ages
based on both their H$\alpha$ morphology and their spectral energy
distribution (SED) fitting were made and show fairly good
correlations.  We also find that the regions with ages determined
younger than 10 Myr are located preferentially along the active
star-forming regions on the spiral arm.
\end{abstract}

\section{Introduction}
Detailed studies of resolved stellar populations allow us to study
stellar evolution and star formation history of galaxies in
detail.  With the construction of large ground-based telescopes with
sensitive images and the launch of the HST, the observational
astronomers in the field of stellar population studies have been able
to expand their observable universe from the nearby clusters in the
Milky Way and Magellanic Clouds to the clusters in extragalactic
space and have also started to compare stellar evolutionary theories to
the observations with unprecedented depth and accuracy.  Since the H-R
diagram was introduced to the astronomical community in
1910, it has been one of the most frequently used plots and
understanding the correlation between the color, magnitude, and the
evolutionary phase of stars is one of the most fundamental subject in stellar
astrophysics.\\ 
M83, also known as the ``Southern Pinwheel'' galaxy, is an interesting
laboratory to study stellar populations.  It is a massive barred
spiral galaxy located at a distance of 4.5\,Mpc \citep{thim03}
with a starbursting nucleus.  Along the spiral arms, H$\alpha$ regions
in different scales display recently formed massive stars, while you
can find red supergiants throughout the galaxy.  By plotting stellar
evolution models and comparing them to observed color-magnitude
diagrams (CMDs), we are able to determine ages of stars in galaxies.
However, in some of the regions with recent starbursting activity, the
young stellar populations are partially obscured by dust. Therefore,
spatial variations in the amount of dust can cause the over-estimates
of stellar ages from the isochrone fitting in the CMDs if we only
apply a single value of internal extinction to correct for the
individual stars in an entire region.  By using the additional color
information, we can constrain the variations of dust extinction in the
galaxy and make corrections for individual stars. \\
The recently installed Wide Field Camera 3 (WFC3) on the Hubble Space
Telescope allows us to study individual stars in M83 in unprecedented
detail, primarily because of the much higher values of discovery
efficiency (quantum efficiency $\times$ field of view) in WFC3.  The
HST/WFC3 observations of M83 were performed in August in 2009 as part
of the WFC Science Oversight Committee (SOC) Early Release Science
(ERS) program (ID:11360, PI: Robert O'Connell). The central region
(3.2$\times$3.2 kpc$^2$) of M83 was observed during the first
observing run. The second field, the adjacent NNW field, will be
included in future publication. \\

\section{HST/WFC3 Observations and Data Analysis}
The data were obtained in seven broad- and seven narrow-band filters
in the UVIS and IR channel. In this study, we used four broad-band
images obtained in F336W (''U''), F438W (''B''), F555W (''V''), and
F814W (''I''). Details about the data are described in
\citet{dopita10} and \citet{chandar10}. The raw data were processed
using the MultiDrizzle task \citep{koekemoer02} with the effective
pixel size of 0.04$\arcsec$ which corresponds to $\sim$0.8 pc per
pixel at a distance of 4.5\,Mpc \citep{thim03}. A color composite of
F336W$+$F555W$+$F814W (broad bands) and F502N$+$F657N (narrow bands)
images is shown in Figure 1. It covers the nuclear region, part of the
spiral arm, and the inter-arm region. PSF photometric analysis of the
WFC3 data was performed using DoPhot \citep{schechter93}. We selected
50 regions indicated in boxes with region numbers to study the spatial
distribution of young stellar populations in M83 and compare their
ages to the cluster ages in the same region.\\ 
\noindent
{\bf Extinction Correction :} The foreground Galactic extinctions for
M83 are 0.361 ($A_{F336W}$), 0.290 ($A_{F438W}$), 0.229 ($A_{F555W}$),
and 0.133 ($A_{F814W}$) mag calculated for the WFC3 UVIS filters using
the extinction curve of \citet{schlegel98}. As we mentioned earlier in
this paper, since M83 has an intricate structure of dust lanes
associated with active star forming regions in the spiral arms and
thin layers of dust in the inter-arm area, we cannot use a single
value of internal extinction and apply it to correct the extinction
for all stars detected in a region of our images.  Therefore, we used the
traditional method of estimating intrinsic colors of individual stars
from the reddening free parameter $Q$:
\begin{equation}
Q_{UBVI} = (U-B) - \frac{E(U-B)}{E(V-I)} \times (V-I) .
\end{equation}
First, we use the slope of the reddening curve for the Milky Way, 0.58
for $E(U-B)/E(V-I)$, derived from the measured $UBVI$ photometry of
early type stars \citep{massey95}.  We also determine the empirical
slope of the reddening vector in the (U-B) vs. (V-I) diagram for
certain regions where the vector is well defined in our own data
(e.g. color-color diagram of region 4 in Figure 2) and apply it to
calculate Q values of the Padova stellar isochrones and the
observational data.  We then match the observed data point to its
corresponding intrinsic value from the Padova model using the Q
value. We do this for both the classic slope of 0.58 appropriate for the
Milky Way and our own empirically determined values based on our M83
data.  In general the resulting corrections are only slightly
different.  Corrected data points are relatively well matched with the
stellar isochrones. \\
\noindent 
{\bf Stellar Ages from Isochrone Fitting :} Fifty regions were
selected in the spiral arm and the inter-arm area to determine the
recent star formation history of M83 (see Figure 1).  Each region was
classified based on their H$\alpha$ morphology (see \citet{whitmore10}
for detail).  The age of the dominant stellar population in each
region was determined by isochrone-fitting. Age-estimates in 50
regions range from 1 Myr to 30 Myr.  Only sources detected with a
signal-to-noise ratio larger than $\sim$3 were used to plot the CMDs
in Figure 2. In regions classified as (4a), young and blue luminous
stars are moderately reddened ($A_{V}$ = $\sim$0.5- 1.0) by dust.  The
39 regions that show evidence of recent star formation range in age
from 1 Myr to 10 Myr. They are located in the ``Pearls'' in the spiral
arm \citep{toomre77}. \\

\section{Summary and Future Work}
The CMD and color-color diagrams of resolved stars from the multi-band
HST/WFC3 observations of M83 indicate the presence of multiple stellar
population features, including the recently formed MS, possible
He-burning blue-loop stars, RGB, and AGB stars. The regions with ages
determined younger than 10 Myr are located along the active
star-forming region on the spiral arm.  Ages of young stellar
populations in each selected region are in fairly good agreement with
the cluster age-datings from SED fitting and from H$\alpha$
morphology.  Near-IR data (F110W and F160W) and the adjacent NNW field
will be added to the analysis in future. \\

\begin{figure}
\center
\includegraphics[scale=0.20,keepaspectratio=yes]{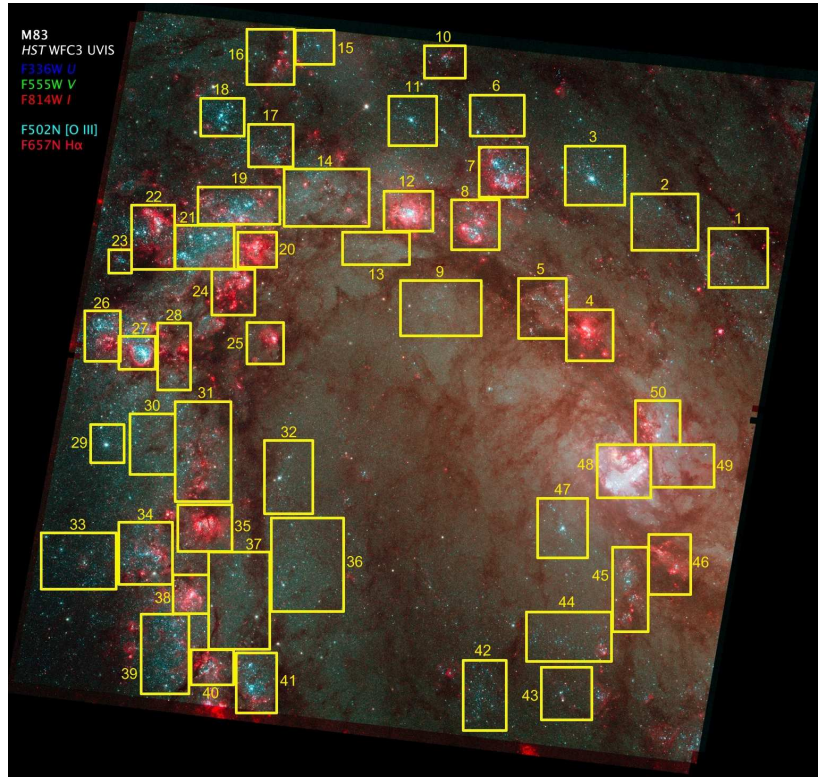}
%\plotone{m83_reg50.eps}
\caption{Color composite of F336W+F555W+F814W (broad bands) and
  F502N+F657N (narrow bands) images with 50 selected regions
  overlaid.
\label{fig:reg50}
}
\end{figure}

\begin{figure}
\center
\includegraphics[scale=0.2,keepaspectratio=yes]{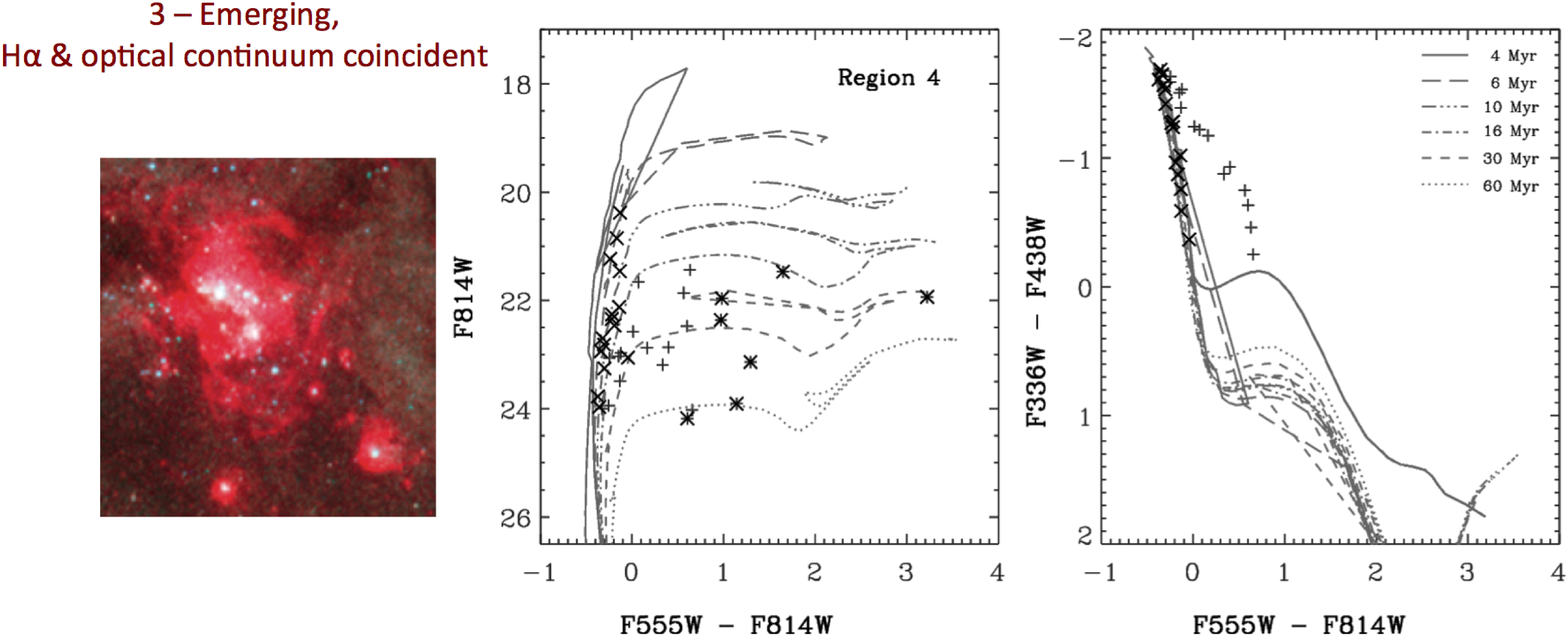}
\includegraphics[scale=0.2,keepaspectratio=yes]{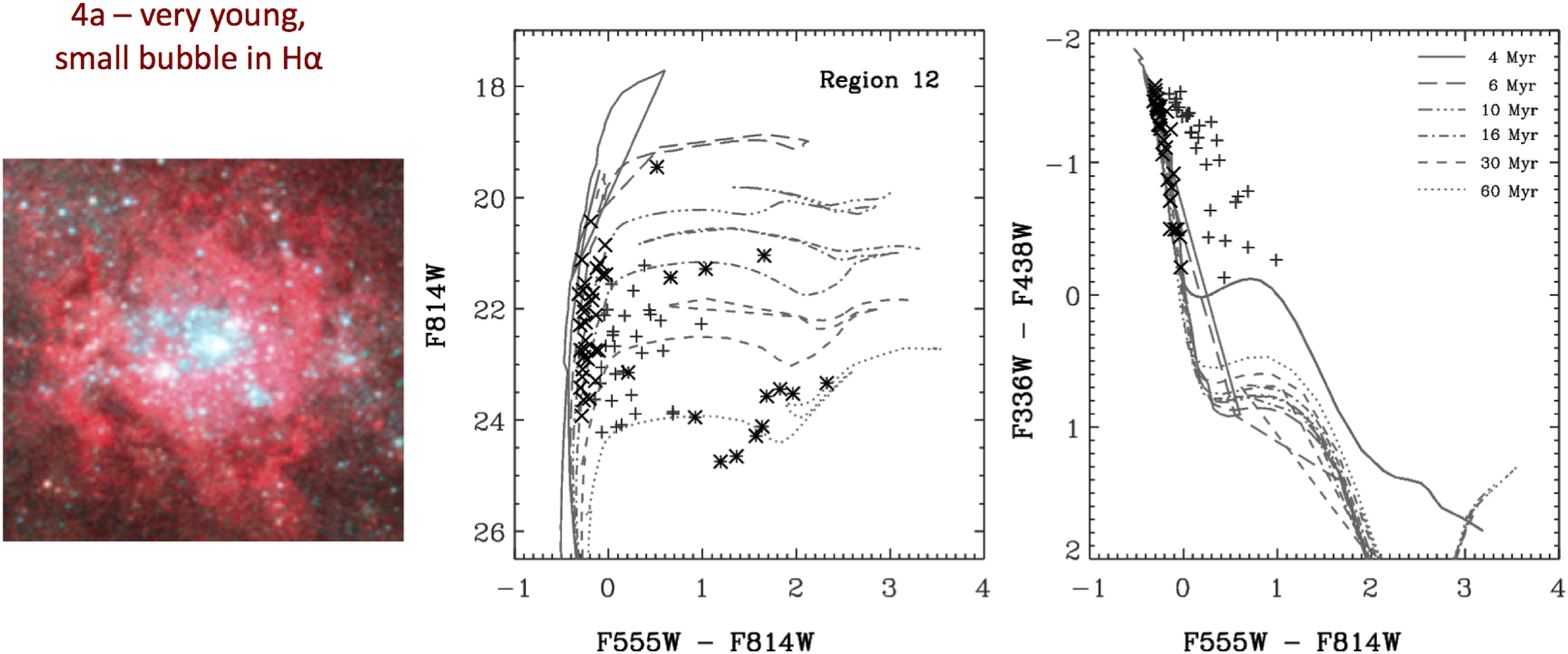}
\includegraphics[scale=0.2,keepaspectratio=yes]{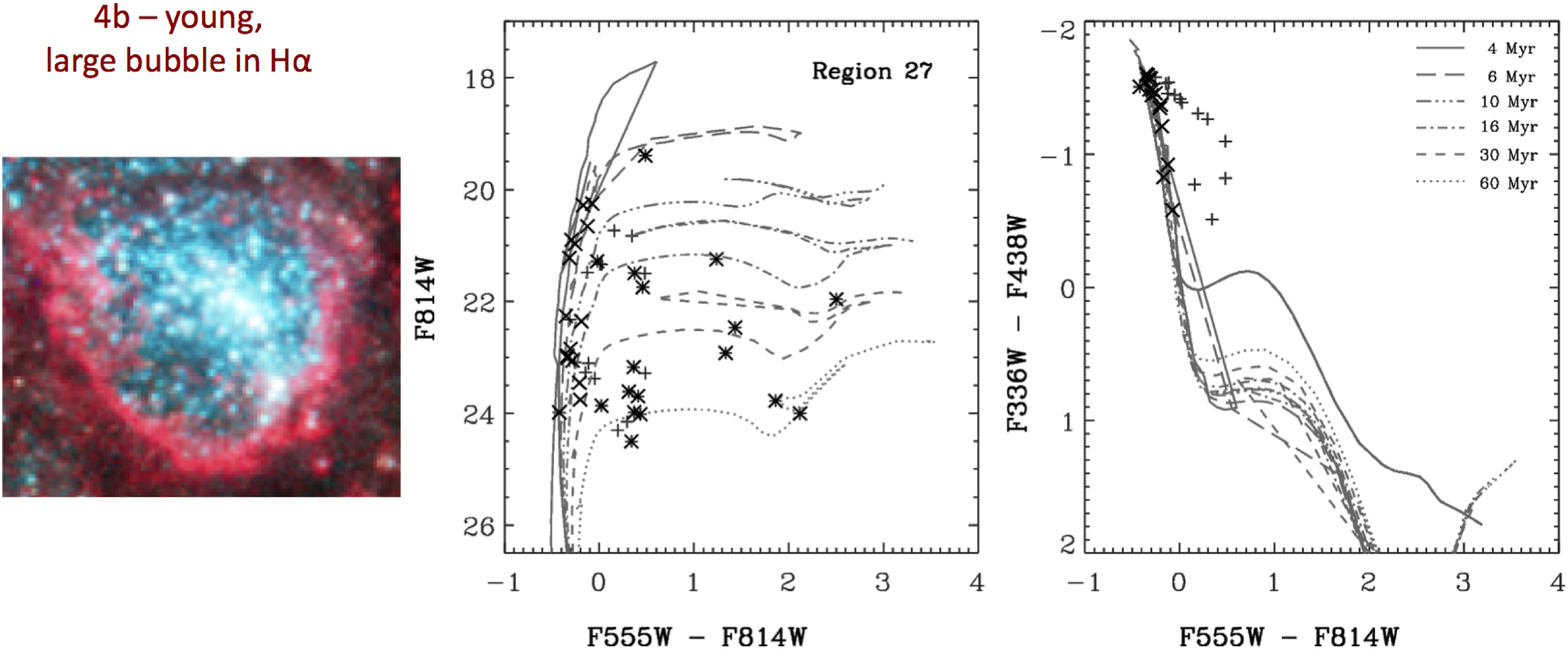}
\includegraphics[scale=0.2,keepaspectratio=yes]{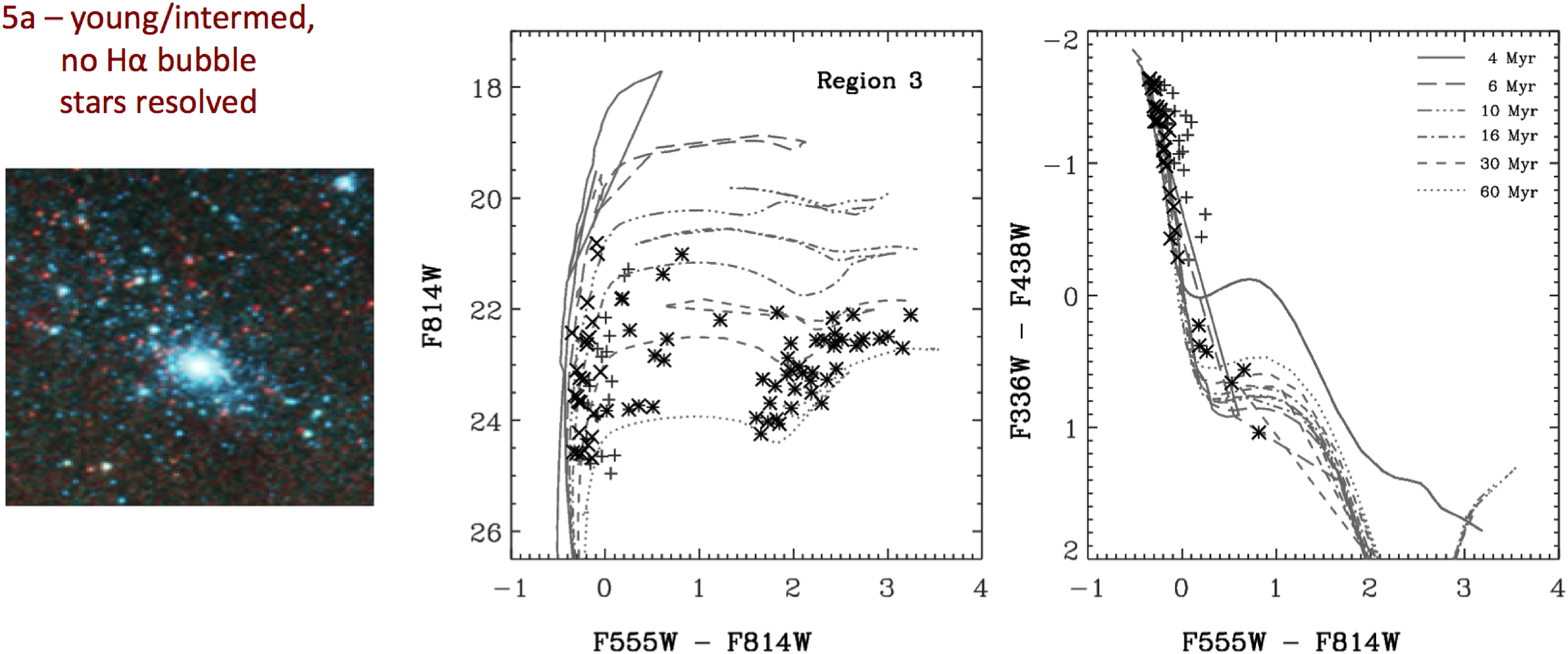}
\caption{CMDs and color-color diagrams. The region classification for
  each category of H$\alpha$ morphology is described above the image
  cut-outs. Data points before and after extinction correction are
  plotted as crosses ``+'' and asterisks ``*''.  Padova isochrones for
  Z=0.03 (1.5 Z$\odot$) and ages of 4, 6, 10, 16, 30, \& 60 Myr are
  overlaid in the CMDs and color-color diagrams.
\label{fig:agedating}
}
\end{figure}

\acknowledgements 
We are grateful to the Director of the Space Telescope Science
Institute for awarding Director's Discretionary time for this program.
Support for program \#11360 was provided by NASA through a grant
from the Space Telescope Science Institute, which is operated by the
Association of Universities for Research in Astronomy, Inc., under
NASA contract NAS 5-26555.

%\bibliography{up2010_m83_star_ref}
\bibliography{up2010_refs_m83.tex}

\end{document}